\documentclass[conference, 9pt, a4paper]{IEEEtran}

\usepackage[utf8]{inputenc}
\usepackage{amssymb}
\usepackage{amsmath}
\usepackage{graphicx}
\usepackage{cite}
\usepackage{enumitem}
\usepackage{url}
\usepackage{booktabs}
\usepackage{multirow}
\usepackage{hyperref}
\usepackage[caption=false]{subfig}
\usepackage{algorithm}
\usepackage{algpseudocode}
\usepackage{balance}

\begin{document}

\title{PATCH-FFT: Unmasking Dormant Hardware Trojans with Patch-Based Frequency-Domain Transformers}

\author{\IEEEauthorblockN{Hasala Senevirathne}
\IEEEauthorblockA{\textit{Computer Engineering \& Computer Science Department} \\
\textit{California State University Long Beach}\\
Long Beach, CA, USA \\
hasala.senevirathne01@student.csulb.edu}
\and
\IEEEauthorblockN{Amin Rezaei}
\IEEEauthorblockA{\textit{Computer Engineering \& Computer Science Department} \\
\textit{California State University Long Beach}\\
Long Beach, CA, USA \\
amin.rezaei@csulb.edu}
}

\maketitle

\begin{abstract}
Hardware Trojans embedded by malicious entities in integrated circuits can covertly leak sensitive information through power side channels, often remaining undetected in their dormant state until specific trigger conditions activate their malicious behavior. For information-leaking Trojans, detection in the dormant state is critical, as once triggered, the secret data is already exfiltrated. This paper introduces a patch-based Transformer architecture for detecting dormant hardware Trojans through frequency-domain analysis of power traces. Our approach converts time-domain power measurements into frequency-domain representations using real Fast Fourier Transform (rFFT), revealing spectral signatures hidden in conventional time-series analysis. Experimental results demonstrate that our method achieves 90.94\% average detection accuracy across dormant and active Trojan scenarios, outperforming state-of-the-art approaches particularly in detecting dormant Trojans that prior time-domain methods do not address. \\
\end{abstract}


\section{Introduction}
Globalized semiconductor supply chains have introduced serious security risks in modern Integrated Circuits (ICs), with Hardware Trojans (HTs) among the most dangerous~\cite{tehranipoor2010survey,karri2010survey}. These malicious modifications can leak cryptographic keys, steal data, or undermine system integrity while evading standard tests. Among the most critical class of HTs are information-leaking Trojans, which exfiltrate secret data through covert side channels. For these Trojans, detection after triggering is too late since the sensitive information has already been compromised. Thus, the only meaningful defense window is detecting the Trojan before it activates.

Contemporary detection approaches predominantly rely on time-domain analysis of side-channel signals, employing Machine Learning (ML) techniques such as Long Short-Term Memory (LSTM)~\cite{nasr2024siamese}, Hierarchical Temporal Memory (HTM)~\cite{Faezi2021HTDetection}, and Convolutional Neural Network (CNN)~\cite{pungati2023multiresolution} to identify anomalous power consumption patterns indicative of ``active'' HTs. Static-power and IDDQ (quiescent supply current)-based techniques~\cite{wei2012scalable,aarestad2010detecting,lecomte2017onchip} can in principle target dormant Trojans by observing the leakage current contributed by the inserted gates, but they generally require multi-pad probing, embedded on-die sensors, or population-level calibration. Time-domain dynamic-power ML detectors exhibit fundamental limitations when confronting ``dormant'' HTs that induce minimal perturbations in power traces~\cite{Hepp2022HTDetection, Pagliarini2022HTDetection, Vishwakarma2023HTDetection, wang2023secure, Fujimoto2023HTDetection, Vishwakarma2024HTDetection, Vishwakarma2025HTDetection, Shiomi2025HTDetection, Maynard2024HDMitigation, Faruque2025DSHTDetection, Hoque2025HTLLMDetection, Yang2025HTDetection, Hong2026HTDetection}, and only a few of these prior dynamic-power ML works evaluate detection performance in the dormant state~\cite{Senevirathne2026NDCE}.

A critical observation motivating our work is that information-leaking HTs, even when dormant, introduce physical circuit modifications (e.g., modulation logic, shift registers, additional transistor paths) that perturb the power spectrum in structured ways. Modulation-based Trojans, such as those using Code Division Multiple Access (CDMA) encoding, contain clock-driven sequential elements that produce periodic spectral artifacts at specific harmonics. Leakage-based Trojans introduce subtler broadband perturbations through additional sub-threshold current paths. These spectral signatures, while imperceptible in raw time-series data, become evident when power traces are transformed into frequency representations through real Fast Fourier Transform (rFFT) analysis. Moreover, these signatures are localized in frequency bands yet exhibit harmonic relationships across the spectrum, requiring a model that captures both local spectral features and global cross-band dependencies.

Patch-based Transformer architectures~\cite{vaswani2017attention,dosovitskiy2020vit} are naturally suited to this structure: dividing the frequency spectrum into patches captures local spectral coherence, while self-attention over the full patch sequence models the global harmonic relationships characteristic of Trojan-induced modulations. To the best of our knowledge, \textbf{this work is the first patch-based Transformer applied to frequency-domain dynamic-power traces for dormant Trojan detection, requiring neither multi-pad probing nor on-die sensors at inference time}. The key contributions of this work are:
\begin{itemize}[leftmargin=*,itemsep=1pt]
\item Proposing a patch-based Transformer for frequency-domain power traces via segmented spectral analysis, enabling local feature extraction and global context through self-attention;
\item Demonstrating that rFFT-based frequency-domain representations reveal dormant Trojan signatures invisible to time-domain dynamic-power approaches, especially for information-leaking HTs that use modulation-based covert channels;
\item Providing interpretable layer-wise attention maps that reveal which frequency bands drive detection decisions, showing alignment with known Trojan leakage mechanisms;
\item Validating the approach through extensive experiments on eight information-leaking Trojan variants, achieving 90.94\% average accuracy across dormant and active scenarios and remaining stable under synthetic process-variation augmentation.
\end{itemize}

\begin{figure*}[t]
    \centering
    \includegraphics[width=0.92\textwidth]{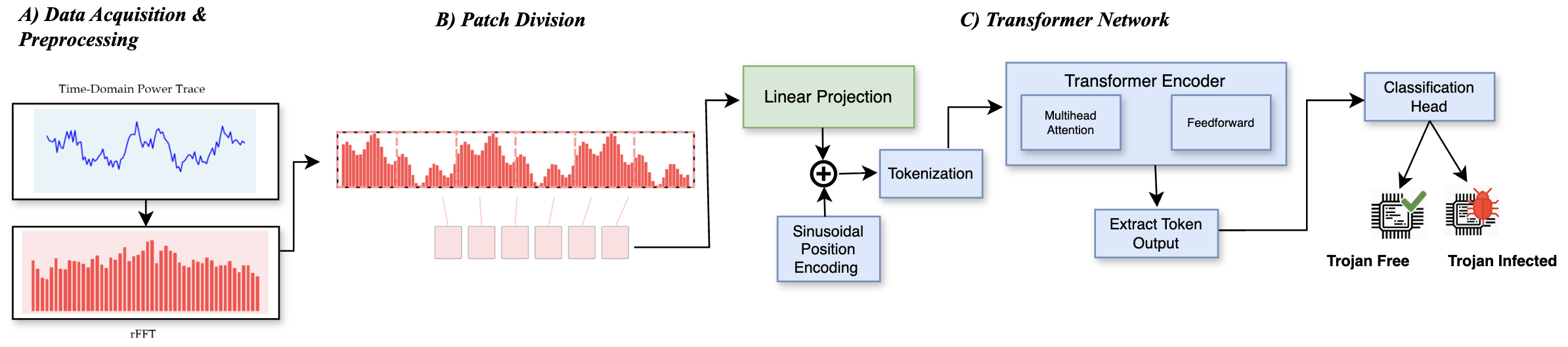}
    \caption{Overview of \textbf{PATCH-FFT}: rFFT log-magnitude $\rightarrow$ 1D patch embedding $\rightarrow$ Transformer encoder $\rightarrow$ binary head.}
    \label{fig:blockdiagram}
\end{figure*}

\section{Related Work}
\textbf{Side-channel HT detection.} Side-channel analysis has proven effective for HT detection. Path delay fingerprinting~\cite{jin2008hardware} identifies timing anomalies but requires extensive characterization. Power consumption analysis~\cite{rad2009sensitivity} compares measurements with golden models but struggles with process variations. In particular, a Siamese deep learning framework is proposed that processes power side-channel traces, reporting up to $86.78\%$ detection accuracy on the Trojan Power \& EM Side-Channel dataset~\cite{nasr2024siamese}. A brain-inspired model known as HTM has also been proposed for HT detection, reporting $92.2\%$ on triggered Trojans only~\cite{Faezi2021HTDetection}. Gate-level netlist approaches such as MLNNs (MLNNs)~\cite{hasegawa2017hardware} and recurrent netlist detectors~\cite{li2019htrnn} form a separate family that does not operate on power traces. None of the above dynamic-power ML works report dormant-state accuracy.

\textbf{Static-power and IDDQ-based detection of dormant Trojans.} A separate line of work targets dormant Trojans by exploiting the leakage current contributed by the inserted gates. Gate-level leakage characterization is formulated as a linear program against the design's gate model, achieving high simulated accuracy but relying on segmentation that becomes ill-conditioned under realistic process variation~\cite{wei2012scalable}. IDDQ is measured simultaneously from multiple supply pads, and spatial anomalies are exploited to localize Trojans, but this approach requires multi-pad probing and population-level calibration~\cite{aarestad2010detecting}. An on-chip $V_{DD}$ sensor array is embedded, and Trojan-induced IR drops are statistically fingerprinted across a lot of dies~\cite{lecomte2017onchip}. While these techniques demonstrate that the dormant-detection problem is not categorically intractable, they require either simulation-only flows, multi-pad/embedded-sensor instrumentation, or golden population statistics. Our work is complementary: we use a single, externally observed dynamic-power trace and a learned spectral model, addressing the same dormant HT detection goal under significantly weaker measurement assumptions.

\textbf{Time series spectral analysis.} FFT-based side-channel methods~\cite{mateos2010correlation,he2014novel} have demonstrated superior noise resilience and feature extraction capabilities compared to time-domain approaches. The rFFT variant leverages the Hermitian symmetry inherent in real-valued signals, offering computational efficiency while retaining all relevant spectral information. The patch-based paradigm, pioneered by Vision Transformers (ViT)~\cite{dosovitskiy2020vit}, treats input data as sequences of local regions while maintaining global context through attention. PatchTST~\cite{nie2023time} demonstrated that patch-based processing of time-series data improves both computational efficiency and modeling capacity. We extend frequency-domain power analysis, where patches map to spectral bands carrying HT-induced modulations.

\section{Architecture}\label{sec:architecture}
In this section, we propose \textbf{PATCH-FFT}, a \textbf{P}atch-based \textbf{A}pproach for \textbf{T}rojan \textbf{C}ountermeasures in \textbf{H}ardware based on \textbf{F}requency-domain \textbf{F}eatures and \textbf{T}ransformers shown in Fig.~\ref{fig:blockdiagram}.  \textbf{PATCH-FFT} identifies three scenarios: (1) \textbf{No Trojan}: The circuit is clean and behaves according to specifications; (2) \textbf{Dormant Trojan}: A Trojan exists but has not yet been triggered; and (3) \textbf{Active Trojan}: A Trojan is triggered and executes its payload. The architecture leverages rFFT to reveal spectral signatures invisible in time-domain representations, followed by patch-based Transformer processing that captures both local frequency features and global patterns of Trojan activity.

\subsection{Frequency-Domain Transformation}
Instead of operating directly on raw time-domain samples, we represent each power trace as a compact frequency-domain signature. Given a discrete-time signal $x[n]$, $n=0,\dots,T{-}1$, we compute the rFFT and retain the $\lfloor T/2 \rfloor{+}1$ non-redundant positive-frequency coefficients exploiting Hermitian symmetry. To obtain a real-valued spectral representation, we map complex coefficients to their magnitudes and apply log-magnitude scaling with per-trace $z$-score normalization:
\begin{equation}
\tilde{X}_{\log}(k) = \frac{\log(|X_{\text{rFFT}}(k)|+\varepsilon) - \mu_x}{\sigma_x + \varepsilon},
\label{eq:logmag}
\end{equation}
where $\mu_x$ and $\sigma_x$ are the mean and standard deviation of $\log(|X_{\text{rFFT}}(\cdot)|+\varepsilon)$ across the $\lfloor T/2 \rfloor{+}1$ frequency bins of trace $x$, and $\varepsilon$ is a small constant for numerical stability. This normalization removes global scale and offset, encouraging the downstream model to focus on relative spectral shapes rather than absolute power levels.

\subsection{Patch Embedding, Transformer, and Classification Head}
The spectrum $\tilde{X}_{\log}\in\mathbb{R}^{F}$ (with $F{=}1251$ for $T{=}2500$) is divided into $N{=}\lceil F/p \rceil = 79$ non-overlapping patches of length $p{=}16$. The final patch is zero-padded to equal length. Each patch is linearly projected to a $d_{\text{model}}{=}128$ token. We add sinusoidal positional encodings~\cite{vaswani2017attention} and prepend a learnable CLS token, producing an 80-token input sequence.

The sequence passes through $L{=}4$ Transformer encoder blocks. Each block applies $h{=}8$-head self-attention:
\begin{equation}
\mathrm{Attention}(Q,K,V)=\mathrm{Softmax}(QK^\top/\sqrt{d_k})V,
\label{eq:attn}
\end{equation}
followed by a $d_{\text{ff}}{=}256$ feed-forward network with residual connections, layer normalization, and dropout $0.1$. The normalized CLS representation is mapped through a linear classification head to two output logits corresponding to \emph{TrojanDisabled} and \emph{TrojanEnabled}. During training, we apply a cross-entropy loss with label smoothing. Algorithm~\ref{alg:detection} summarizes the inference pipeline.

\begin{table}[!t]
\centering
\caption{PATCH-FFT architecture and training hyperparameters.}
\label{tab:hyperparams}
\renewcommand{\arraystretch}{1.0}
\begin{tabular}{|l|c|}
\hline
\textbf{Hyperparameter} & \textbf{Value} \\
\hline\hline
Input length $F = \lfloor T/2 \rfloor + 1$ (with $T = 2500$) & 1251 \\
Patch size $p$                                & 16 \\
Model dimension $d_{\text{model}}$            & 128 \\
Attention heads $h$ ($d_k = d_v = 16$)        & 8 \\
Encoder layers $L$                            & 4 \\
FFN dimension $d_{\text{ff}}$                 & 256 \\
Dropout                                       & 0.1 \\
Label-smoothing $\alpha$                      & 0.02 \\
Optimizer / weight decay                      & AdamW / $10^{-4}$ \\
Learning rate (cosine, 2-epoch warm-up)       & $10^{-3}$ \\
Batch size                                    & 128 \\
\hline
\end{tabular}
\end{table}

\begin{algorithm}[t]
\caption{PATCH-FFT inference for one power trace. The same procedure is invoked with the dormant-detection checkpoint (positive class = \emph{TrojanEnabled}) or the active-detection checkpoint (positive class = \emph{TrojanTriggered}).}
\label{alg:detection}
\begin{algorithmic}[1]
\Function{DetectTrojan}{$x$, model, positive\_class}
    \State $X \gets \mathrm{rFFT}(x)$
    \State $X_{\log} \gets \log(|X| + \varepsilon)$
    \State $\tilde{X}_{\log} \gets z\text{-score}(X_{\log})$ \Comment{$z$-score over frequency bins}
    \State $\mathbf{h}_{\mathrm{cls}} \gets \text{TransformerEncoder}(\text{Patchify}(\tilde{X}_{\log}))$
    \State $\hat{y} \gets \arg\max\bigl(\mathrm{Softmax}(W_{\mathrm{cls}}\mathbf{h}_{\mathrm{cls}} + \mathbf{b}_{\mathrm{cls}})\bigr)$
    \If{$\hat{y} = 1$} \Return positive\_class
    \Else \ \Return \emph{TrojanDisabled}
    \EndIf
\EndFunction
\end{algorithmic}
\end{algorithm}

\section{Experimental Setup}\label{sec:setup}

\subsection{Threat Model and Assumptions}
This work targets key-leakage HTs inserted during the design or fabrication process. The evaluation uses the public IEEE/TrustHub dataset, which reproduces this threat on a commercial FPGA platform; claims are scoped to this FPGA-based evaluation, and ASIC silicon validation is left as future work.

\textbf{PATCH-FFT} requires golden chips only during training, where labeled Trojan-free and Trojan-induced traces are used to learn the classifier. At deployment, the model takes a single externally measured power trace from the device under test and outputs a binary verdict. The measurement requires only an external shunt resistor and oscilloscope, with no multi-pad probing, on-die sensors, IDDQ instrumentation, or invasive optical access, matching golden-free inference~\cite{Faezi2021HTDetection, nasr2024siamese} and avoiding extra instrumentation~\cite{aarestad2010detecting,lecomte2017onchip}.

\subsection{Dataset and Preprocessing}
We evaluate \textbf{PATCH-FFT} using the publicly available IEEE HT Power \& EM Side-Channel Dataset~\cite{dataset_ieee}, which includes power traces from AES-128 implementations carrying TrustHub Trojan variants. The target design is synthesized onto the Xilinx Spartan-6 (45\,nm) cryptographic FPGA of a SAKURA-G side-channel evaluation board~\cite{satoh2014sakura}; the supply current is acquired through the on-board shunt resistor on the $V_{cc\text{-}int}$ rail and digitized by an external oscilloscope. All traces are captured at $25^\circ\mathrm{C}$ from a single FPGA instance per benchmark, so the public dataset does not span natural die-to-die process variation. For each Trojan configuration (Trojan-free baseline, Trojan-disabled/dormant, Trojan-triggered/active) the dataset provides on the order of $10{,}000$ traces, where each trace is a univariate time series of length $T = 2500$ samples captured during one AES encryption. For this study, we focus on eight Trojan variants that employ key-leakage mechanisms. Although AES-T400/T1600/T1700 use ``RF/EM radiation'' as the listed leakage channel, the modulator and counter logic that drives those off-die emissions also draws clock-correlated current from the FPGA core rail; the supply-rail shunt trace carries a weaker replica of the same modulation, and we retain it as the sole input for all eight benchmarks.

\subsection{Training Setup}\label{subsec:split}
We use the power side-channel traces and reorganize them into two binary classes, \emph{TrojanDisabled} and \emph{TrojanEnabled} or \emph{TrojanDisabled} and \emph{TrojanTriggered}. For each AES-Tx benchmark we (i) pool all available traces belonging to the two relevant classes, (ii) shuffle them with a fixed seed, and (iii) partition them into a $70\%/10\%/20\%$ training / validation / test split, stratified by class label, yielding approximately $14{,}000$ / $2{,}000$ / $4{,}000$ traces per benchmark. The three subsets are disjoint at the trace level; no trace appears in more than one subset, and a separate model checkpoint is trained per benchmark. The dormant detection accuracy is computed on the held-out $20\%$ test split, and the active accuracy is obtained analogously by replacing the positive class with \emph{TrojanTriggered}. This is the same intra-benchmark evaluation protocol adopted by~\cite{nasr2024siamese} and \cite{Faezi2021HTDetection}; cross-benchmark transfer is left to future work.

The model is trained end-to-end using cross-entropy loss with label smoothing ($\alpha = 0.02$), AdamW optimizer ($\text{lr}{=}10^{-3}$) with cosine annealing schedule and linear warmup for the first 2 epochs, batch size 128, and early stopping with patience of 5 epochs based on validation accuracy.

\section{Experimental Results}\label{sec:results}

\begin{table*}[t]
\centering
\caption{PATCH-FFT accuracy across different AES hardware Trojan variants.}
\label{tab:accuracy_result}
\renewcommand{\arraystretch}{1.0}
\begin{tabular}{|p{1.4cm}|p{6.6cm}|p{3.3cm}|p{2.2cm}|p{2.0cm}|}
\hline
\textbf{Benchmark} & \textbf{Key-Leakage Mechanism} & \textbf{Native Channel} & \textbf{Dormant (\%)} & \textbf{Active (\%)} \\
\hline\hline
AES-T600  & Leakage current modulated by key bits                      & Leakage current  & 77.3 & 83.6 \\\hline
AES-T2000 & Increased leakage current for key bits equal to 0          & Leakage current  & 88.4 & 92.1 \\\hline
AES-T700  & CDMA-based modulation of dynamic power after trigger seq.  & Dynamic power    & 96.7 & 99.2 \\\hline
AES-T800  & CDMA-based modulation of dynamic power after trigger seq.  & Dynamic power    & 100.0 & 100.0 \\\hline
AES-T1000 & Increased dynamic power after trigger sequence             & Dynamic power    & 95.5 & 97.8 \\\hline
AES-T400  & Key bits transmitted via modulated RF signal on unused pin & RF / EM emission & 89.2 & 92.4 \\\hline
AES-T1600 & Key bits transmitted via RF signal after trigger sequence  & RF / EM emission & 82.8 & 84.3 \\\hline
AES-T1700 & Key bits transmitted via RF signal driven by counter       & RF / EM emission & 86.6 & 89.1 \\\hline
\end{tabular}
\end{table*}

\subsection{Frequency-Domain Signatures of HTs}
\begin{figure}[!t]
\centering
\includegraphics[width=0.95\columnwidth]{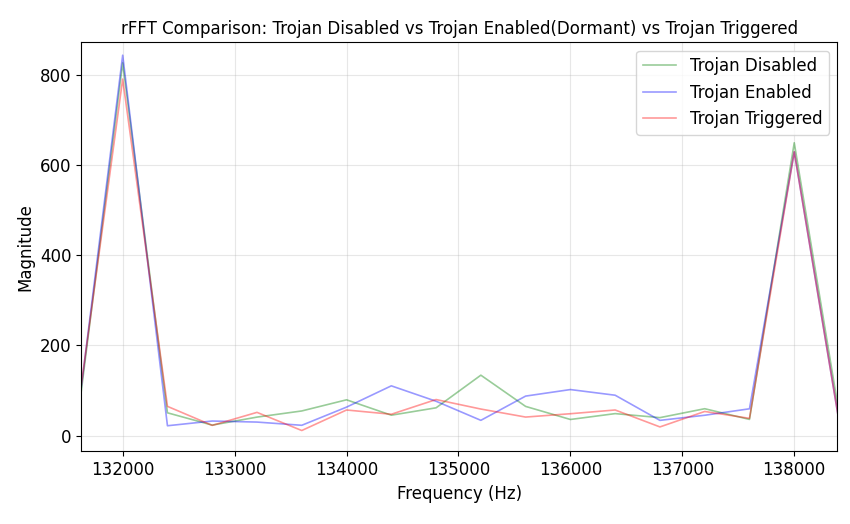}
\caption{rFFT magnitude spectrum of AES-T400 in baseline, dormant, and active modes, zoomed to the band where the three modes differ most (dataset-native frequency units; see Sec.~\ref{sec:setup}).}
\label{fig:fft_comparison}
\end{figure}

Fig.~\ref{fig:fft_comparison} compares rFFT magnitudes for a representative AES-T400 trace in baseline, dormant, and active modes. Individual bin differences are too subtle to isolate by inspection. However, they become discriminative when aggregated over adjacent non overlapping frequency patches with global self attention, a property \textbf{PATCH-FFT} exploits.

\subsection{Performance Analysis}
Table~\ref{tab:accuracy_result} reports detection accuracy across eight key-leakage HTs. \textbf{PATCH-FFT} achieves $89.56\%$ dormant and $92.31\%$ active accuracy ($90.94\%$ average). CDMA-modulation Trojans (T700/T800/T1000) include free-running counter and PN-sequence logic on the AES clock. This logic switches even when the payload trigger has not fired, producing sharp spectral peaks that the model detects with near-perfect accuracy. RF-based Trojans (T400/T1600/T1700) produce moderate spectral artifacts. Leakage-current Trojans (T600/T2000) create only broadband sub-threshold perturbations, making them the hardest variants to separate from PV noise.

\subsection{Process-Variation Robustness}\label{subsec:pv}
Because the dataset contains a single FPGA instance, we synthesize PV replicas by perturbing each trace as
\begin{equation}
x'[n] = (1+\alpha)\, x[n] + \eta_{\mathrm{LP}}[n],\quad \alpha\sim\mathcal{N}(0,\sigma^2),
\label{eq:pvmodel}
\end{equation}
with $\eta_{\mathrm{LP}}$ zero-mean Gaussian noise low-pass filtered to the FPGA core bandwidth and RMS-scaled by $\sigma\!\cdot\!\mathrm{RMS}(x)$. The multiplicative term captures die-to-die $V_{th}/I_{dsat}$ shifts. The band-limited additive term emulates correlated within-die variability~\cite{yuan2021pvresistant,sun2022em,bernstein2006variability}. Training data is clean; perturbation is applied only to the held-out test split (10 seeds per $\sigma$).

Table~\ref{tab:pv_ablation} shows that dormant accuracy degrades by $<\!1$\,pp at $\sigma{=}5\%$ and retains $>\!84\%$ at $\sigma{=}15\%$. An ablation isolating each term reveals that $z$-score normalization absorbs the multiplicative gain. The additive band-limited noise drives nearly all of the degradation. This confirms that \textbf{PATCH-FFT}'s spectral-locality property tolerates die-mean shifts.

\begin{table}[!t]
\centering
\caption{PATCH-FFT dormant-state accuracy under synthetic process variation (mean over eight benchmarks, ten seeds).}
\label{tab:pv_ablation}
\begin{tabular}{|c|c|c|}
\hline
\textbf{$\sigma$ (\% RMS)} & \textbf{Dormant Acc.\ (\%)} & \textbf{$\Delta$ vs.\ clean} \\
\hline\hline
0  & 89.56 & --      \\
5  & 88.71 & $-0.85$ \\
10 & 86.94 & $-2.62$ \\
15 & 84.10 & $-5.46$ \\
20 & 80.42 & $-9.14$ \\
\hline
\end{tabular}
\end{table}

\subsection{Comparison with State-of-the-Art Works}
Table~\ref{tab:method_comparison} compares \textbf{PATCH-FFT} to prior side-channel based HT detectors. Crucially, neither HTM~\cite{Faezi2021HTDetection} nor Siamese LSTM~\cite{nasr2024siamese} evaluate or report dormant-state detection accuracy. \textbf{PATCH-FFT} achieves $89.56\%$ on the harder dormant case and $92.31\%$ on triggered Trojans, showing that frequency-domain patch-based analysis enables reliable dormant HT detection, a capability not addressed by prior time-domain methods.

\begin{table}[!t]
\centering
\caption{PATCH-FFT performance against other models.}
\label{tab:method_comparison}
\begin{tabular}{|l|c|}
\hline
\textbf{Method} & \textbf{Accuracy (\%)} \\
\hline\hline
HTM~\cite{Faezi2021HTDetection}      & 92.2 (triggered only) \\
Siamese LSTM~\cite{nasr2024siamese}  & 86.78 (triggered only) \\
\hline
\textbf{PATCH-FFT (dormant)}         & \textbf{89.56} \\
\textbf{PATCH-FFT (triggered)}       & \textbf{92.31} \\
\textbf{PATCH-FFT (average)}         & \textbf{90.94} \\
\hline
\end{tabular}
\end{table}

\begin{figure}[!t]
\centering
\subfloat[]{\includegraphics[width=0.95\linewidth]{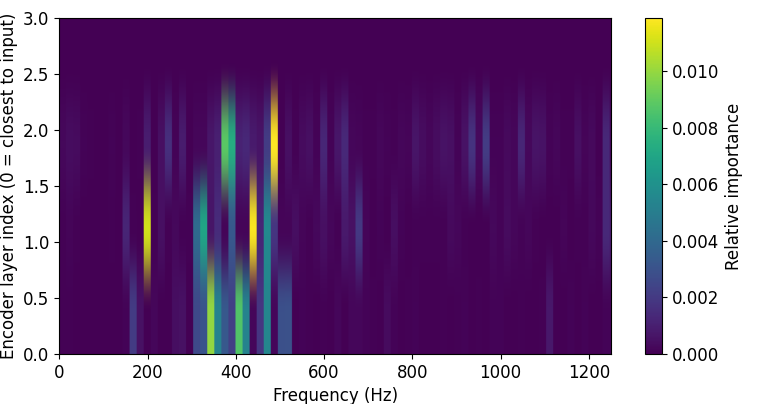}%
\label{fig:heatmap_t600}}
\hfil
\subfloat[]{\includegraphics[width=0.95\linewidth]{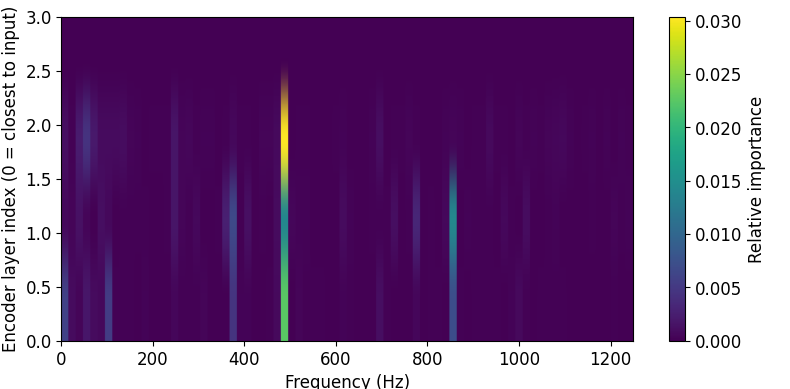}%
\label{fig:heatmap_t800}}
\caption{Attention heatmaps for different Trojan variants: (a) AES-T600 (b) AES-T800.}
\label{fig:attention_heatmaps}
\end{figure}

\subsection{Layer-wise Frequency Sensitivity}
Fig.~\ref{fig:attention_heatmaps} shows encoder layer--frequency heatmaps for AES-T600 and AES-T800. For AES-T600, early layers respond broadly while deeper layers concentrate on narrow bands where dormant traces deviate from the baseline. For AES-T800, attention quickly collapses onto harmonics of the CDMA spreading sequence, confirming that the model learns physically meaningful spectral features rather than spurious correlations.

\section{Conclusion}
In this paper, we presented \textbf{PATCH-FFT}, a patch-based frequency-domain Transformer for HT detection from power side-channel traces. By operating on rFFT spectra and attending to frequency patches, the model achieves strong accuracy not only in detecting active Trojans but also in identifying dormant Trojans, a capability that prior dynamic-power ML methods do not address. Modulation-based HTs that produce sharp spectral signatures are detected with near-perfect accuracy, while subtler leakage-current variants remain more challenging. 

\textbf{PATCH-FFT} achieves $89.56\%$ dormant and $92.31\%$ active accuracy ($90.94\%$ average) and remains stable under synthetic process-variation augmentation. Future work will focus on improving detection of leakage-current Trojans through time- and frequency-domain feature fusion, evaluating cross-benchmark transfer, validating the approach on EM traces, and extending evaluation to fabricated ASIC silicon.


\end{document}